\documentclass{vgtc}                          

\ifpdf
  \pdfoutput=1\relax                   
  \usepackage{graphicx}                
  \DeclareGraphicsExtensions{.pdf,.png,.jpg,.jpeg} 
\else
  \usepackage{graphicx}                
  \DeclareGraphicsExtensions{.eps}     
\fi%

\graphicspath{{figures/}{pictures/}{images/}{./}} 

\usepackage{microtype}                 
\PassOptionsToPackage{warn}{textcomp}  
\usepackage{textcomp}                  
\usepackage{mathptmx}                  
\usepackage{times}                     
\usepackage{cite}                      
\usepackage{tabu}                      
\usepackage{booktabs}                  

\usepackage{xurl}
\usepackage{paralist}
\usepackage{tikz}
\usepackage{cancel}
\def\checkmark{\tikz\fill[scale=0.4](0,.35) -- (.25,0) -- (1,.7) -- (.25,.15) -- cycle;} 
\usepackage{pifont}
\newcommand{\xmark}{\ding{53}}%

\newcommand{\cmmnt}[1]{}

\newcommand{\rv}[1]{\textcolor{black}{#1}}

\onlineid{1143}

\vgtccategory{Theory or Model}

\begin{document}

\title{Toward Systematic Design Considerations of Organizing Multiple Views}





\addtolength{\footnotesep}{-10mm} 

\newcommand*\samethanks[1][\value{footnote}]{\footnotemark[#1]}

\author{Abdul Rahman Shaikh\thanks{e-mail: ashaikh2@niu.edu.} %
\and David Koop\thanks{e-mail: dakoop@niu.edu.}
\and Hamed Alhoori \thanks{e-mail: alhoori@niu.edu.}
\and Maoyuan Sun \thanks{e-mail: smaoyuan@niu.edu.}}
\affiliation{\scriptsize Northern Illinois University}

\teaser{
\centering
 \includegraphics[width=16cm,height=8.9cm]{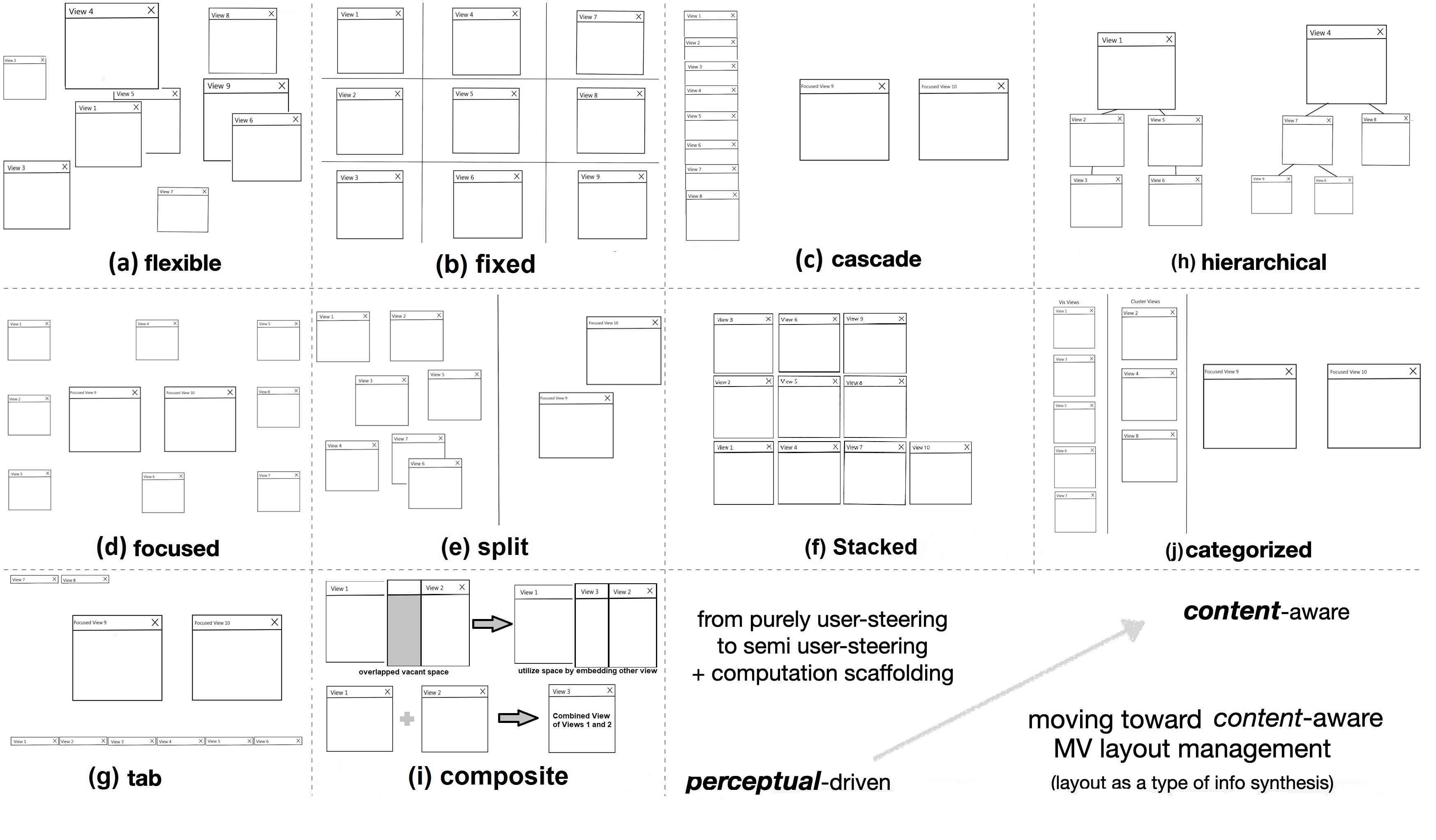}
 \vspace{-6.2mm}
 \caption{\rv{Examples of MV layout designs based on factors of user perception (a-g) and view content (h-j), which highlights broadening the design considerations from purely perception-driven to intelligently content-driven.} 
}
 \label{fig:layout}
}

\abstract{%
Multiple-view visualization (MV) has been used for visual analytics in various fields (e.g., bioinformatics, cybersecurity, and intelligence analysis).
Because each view encodes data from a particular perspective, analysts often use a set of views laid out in 2D space to link and synthesize information.
The difficulty of this process is impacted by the spatial organization of these views.
For instance, connecting information from views far from each other can be more challenging than neighboring ones.
However, most visual analysis tools currently either fix the positions of the views or completely delegate this organization of views to users (who must manually drag and move views).
This either limits user involvement in managing the layout of MV or is overly flexible without much guidance.
Then, a key design challenge in MV layout is determining the factors in a spatial organization that impact understanding.
To address this, we review a set of MV-based systems and identify considerations for MV layout rooted in two key concerns: \textit{perception}, which considers how users perceive view relationships, and \textit{content}, which considers the relationships in the data.
We show how these allow us to study and analyze the design of MV layout systematically.

} 







\firstsection{Introduction}
\maketitle









\textit{Multiple-view visualization} (MV) has been commonly used in visual analysis tools for sensemaking of data in various application domains, including multimedia analysis (e.g., Canopy \cite{burtner2013interactive}), text analytics (e.g., IN-SPIRE \cite{inspire} and Jigsaw \cite{jigsaw}), business intelligence (e.g., Spotfire \cite{ahlberg1996spotfire} and Tableau \cite{tableau}) and cyber security \cite{zhang2015visualizing}.
To support sensemaking activities, users often work with multiple views and spatially organize them, if allowed, on a display space (e.g., placing a focused view in the center, moving a few related ones around, and minimizing unused ones).
Designing effective visualizations or MV systems to explore a dataset effectively is challenging due to the restrictive layouts, complicated coordination mechanisms, and interactions between multiple dimensions of design spaces \cite{10.1145/345513.345271}.

As prior research shows that humans can use space to think \cite{Andrews}, the created layout of MV can encode specific meanings that assist in the information displayed in multiple views. 
However, current visual analysis tools either use a fixed layout of MV \cite{birdvis,hofman} or completely delegate organizing MV to users \cite{jigsaw,similarity} (e.g., manually moving views). 
The former limits user involvement in organizing MV, and the latter seems overly flexible as it does not provide any scaffold to support users in manually organizing MV.
While recent research performed an in-depth study of the composition and configuration of MV \cite{chen2020composition}, their MV layout design recommendation considers little about the content in MV and cross-view data relationships, which impacts the usage of MV \cite{sun2021towards}.

This leads to an important design challenge for MV layout. 
Specifically, what important factors should we consider for the design of MV layout?
To address this, we conducted a systematic review of 360 MV designs used in a prior study \cite{chen2020composition}. 
Based on the review, we propose two major aspects to consider while designing the layout of MV: \textit{perception} and \textit{content}.
The former considers user perception of MV. 
The latter regards data relations across MV.

We argue that a view’s content and the cross-view data relationship should be a key consideration for designing the layout for an MV system.
MV systems should account for view content as layouts tailored to a prominent view can help users be more productive \cite{boger2021jurassic}.
Specifically, well-designed systems in this respect can reduce the time spent exploring the dataset by effectively conveying essential information to users by avoiding clutter in display space \cite{ajani2021declutter}. 
An improved layout is also correlated to the productivity of and interpretability by the user \cite{reppa}.
However, despite this key benefit, view content or cross-view data relations are not extensively considered in existing design techniques and tools for layouts of MV.

This study offers a novel approach to the design layout of MV systems based on user perception and view content. 
We identify factors related to them to build a practical design layout for MV systems. 
We further discuss design layouts in detail based on these attributes and factors. 
Our work suggests layouts to assist designers in planning an initial point in their layout design and consider content as an essential part of their layout design process.
Our contributions in this paper are as follows:
\vspace{-5pt}
\begin{itemize}
    \setlength{\parsep}{0em} 
    \setlength{\itemsep}{0em} 
    \item We describe how user perception and view content are essential attributes to build concrete design guidelines for MVs.
    \item We present layouts that help designers consider perception and content when designing MVs.
\end{itemize}

\section{Related Work}
\subsection{Multiple View}
The field of MV \cite{North01,VisTrails,Ryu2003ExploringCS,MVmeanings} has been explored extensively in the past few years. 
Researchers have focused on encouraging the use of MV \cite{RobertsEnc}, providing guidelines for using MVs \cite{10.1145/345513.345271}, investigating collaborative computational analysis on MVs\cite{Gorg,VisPorter}, exploring multiple coordinated views \cite{Roberts,Scherr2009MultipleAC}
, and creating MV tools or systems to investigate and gain insight into datasets. 
Numerous MV systems 
have been proposed in recent years, including ComVis \cite{comvis}, a tool that facilitates new visualization techniques and prototypes; 
SightBi \cite{Sun_2022}, a visual analytics system that explores cross-view data relationships using biclusters and visualizes them to explore insights into coordinated activities for sensemaking;
JigSaw \cite{jigsaw}, a visual analytic system with MVs containing documents along with entities that can be visually connected to enable analysts to examine the documents quickly and investigate them efficiently. 
\rv{These MV tools and systems have distinct layout designs that provide analysis of data through different perceptions.}

Prior research has analyzed the perception and cognitive performance of users while using different tools \cite{endert2012semantic,endert2014human
}. 
Andrews et al. \cite{Andrews} investigated how increased space affects a user’s cognitive thinking and decision-making processes while performing a sensemaking task. 
They found that factors like design space, design layout, and analytics provenance alter a user’s cognitive thinking, thus affecting their performance in a sensemaking process.
North and Shneiderman \cite{North97ataxonomy} proposed a taxonomy with multiple window coordination capabilities through which users can create visual environments and coordination links between windows.
They found that this coordination between multiple windows was associated with improved user performance, detection of unexpected relationships, and desktop unification (i.e., showing complex information on multiple displays with appropriate coordination).

\subsection{Multiple View Layout}

\rv{Researchers have surveyed the design layout of MV systems and identified a set of different layouts proposed over recent years.}
Chen et al. \cite{chen2020composition} analyzed designs from 360 multiple view systems with different views, layout designs, and display spaces proposed in research articles.
Similarly, Al-maneea and Roberts \cite{maneea} extracted 491 images of MV systems from several conferences, journals, and workshops to explore different view layouts proposed across multiple studies.
However, little to no research has been published that considers the view content or the cross-view data relationship in designing layouts of MV systems.

Several MV systems proposed over the years contain less than five views, simple visualizations, and non-complex layouts often with fixed views \cite{chen2020composition}.
For example, BirdVis \cite{birdvis} is an MV system in which the layout is a side-by-side display of fixed views. 
Hofman et al. \cite{hofman} proposed a system with three fixed views with a central salience view containing a spiral linked view, controls on the left, and a streamgraph view on the right.
The MV system proposed by Bögl et al. \cite{bogl} has four fixed views with two main (salient) views on the top with controls and list views positioned at the bottom.

Apart from simple layouts, a few complex MV systems have been proposed over the last few years, which often use flexible views.
For example, Jigsaw \cite{jigsaw} can accommodate many flexible views in its display space where views can be re-positioned or resized causing overlaps.
Zhao et al. \cite{zhao} proposed a system with flexible resizing of views and a salient view at the center surrounded by its related views. 
The SimilarityExplorer \cite{similarity} system has flexible views similar to Jigsaw but with overlaps that cause users to move views often to analyze two or more focused views simultaneously. 
\rv{In summary, many proposed systems have a fixed or overly flexible layout, as there is no systematic design consideration in organizing MVs.}

\vspace{-5pt}
\section{Design Considerations}
We reviewed a dataset of 360 images of MVs by coding the attributes affecting layout design.
We found the typical design for MV layouts focuses on attributes that affect user perception, often related to the view (e.g., position or size of the view) or viewport (display space) and the content of a view (e.g., visualizations, control panels, and textual data).
We propose that content and perception are essential attributes to consider when designing layouts for MV systems.

We suggest the layout design of an MV system considering the factors of the two coded attributes: \textit{user perception} and \textit{view content}. 
A user’s perception of the view, while working with MVs, affects their cognitive ability to perform an analysis (e.g., a view in a large size or a centered-position can imply it is more important than other views).
\rv{User perception also affects user attention, interpretation, insights, and performance, impacting the decision-making process.}
Similarly, the content present in each view and the relationship between such content also influence a user’s analytical process.
\rv{A view's content is integral to an MV system since it effectively communicates information through appropriate visualizations. The relationship between the content of views provides insightful information and assists users in decision-making.}
\rv{In cross-view data relationships, content in one view may be updated in response to a user action in another view, or elements may be linked between views. (e.g., two views with a cross-view data relationship may convey knowledge more efficiently if they are appropriately positioned near each other rather than separated).}

However, current design processes overlook the cross-view data relationship (relationship between the content of two or more views).
Additionally, we discuss how MV layouts can be improved by integrating cross-view data relationships.
Table \ref{tab:features} provides an overview of the driving factors of each design layout suggested in this work.


\begin{table}[tb]
\centering
\caption{\rv{Proposed Layout Designs (refer to Figure \ref{fig:layout}) and their layout factors categorized based on perception (see Section \ref{Sec3.1}) and content (see Section \ref{Sec3.2}).}
}
\label{tab:features}
\resizebox{\columnwidth}{!}{%
\begin{tabular}{p{1.8cm}|p{0.2cm}|p{0.2cm}|p{0.2cm}|p{0.2cm}|p{0.2cm}|p{0.2cm}|p{0.2cm}|p{0.2cm}|p{0.2cm}|p{0.2cm}}
 \textbf{Factors} & \textbf{(a)} & \textbf{(b)}  & \textbf{(c)} & \textbf{(d)}  & \textbf{(e)}  & \textbf{(f)} & \textbf{(g)} & \textbf{(h)} & \textbf{(i)} & \textbf{(j)} \\
\hline
Separation & \checkmark & \xmark & \checkmark & \checkmark & \checkmark & \checkmark & \checkmark & \checkmark & \checkmark & \checkmark\\\hline
Position & \xmark & \checkmark & \checkmark & \xmark & \xmark & \checkmark & \xmark & \checkmark & \xmark & \checkmark\\\hline
Salience & \xmark & \xmark & \checkmark & \checkmark & \xmark & \xmark & \checkmark & \checkmark & \xmark & \checkmark\\\hline
View content & \xmark & \xmark & \xmark & \xmark & \xmark & \xmark & \xmark & \xmark & \checkmark & \checkmark\\\hline
Cross-view & \xmark & \xmark & \xmark & \xmark & \xmark & \xmark & \xmark & \checkmark & \xmark & \checkmark\\
\end{tabular}
}
\vspace{-0.25in}
\end{table}

\vspace{-0.2cm}
\subsection{Perception}
\label{Sec3.1}

Visual perception is how people process and organize the visual information presented to them; improving a person's perception allows them to efficiently recognize patterns and gain understanding from data~\cite{ware:perception}. 
Perception aids cognition and benefits from encoding information through visual channels like color hue and length. 
In MV layouts, the visual properties of and between the \emph{views} significantly contribute to how this information is understood.
Thus, we define a \emph{perception-driven layout} as a layout that organizes views to improve visual perception and thus reduce cognitive load. 
This can be achieved by differentiating different views, drawing user attention to important views, and helping users navigate through views.


We have categorized a number of factors, including \emph{separation}, \emph{relative position}, and \emph{salience}, which contribute to the perception of MVs.
The most important property of an MV system is \emph{separation}. 
It allows the user to distinguish between two or more views.
This does not mean that views cannot overlap, but if they do, users need to be able to clearly identify individual views. 
 \rv{However, in some MV systems, users may not always need to distinguish between individual views (e.g., if two views are very similar and can provide more information if perceived as one).}
The second factor influencing perception-driven layouts is the \emph{relative position} of the views. 
A set of views that are randomly laid out instead of aligned in a grid affects the way relationships between views are understood.
Similarly, two views located far away from each other are often interpreted as unrelated, and a view in a central position may be understood as a base for those that are eccentric.
The third factor we have identified is \emph{salience} which captures how particular views draw more attention than others.
For example, if most views use a grayscale colormap, but one view uses vivid colors, that view is more salient.
Other properties influencing saliency include whether a view is minimized or maximized or is larger or smaller in size.
These factors intersect as well; a central view that is also larger than others may be identified as the most important view in the visualization.

\vspace{-0.25cm}
\subsection{Content}
\label{Sec3.2}
We define content-aware layouts as those being focused on each view’s content (e.g., types of visualizations and control panels) and the relationship of content between different views.
In content-aware layouts, we focus on the relationship between views to reveal hidden patterns or relationships between content elements.
We categorize the content-aware factor as the \textit{cross-view data relationship} between views.
As the content of the views impacts an MV system's analysis process, we propose content-aware layouts that consider view content and cross-view data relationships when designing layouts.

The cross-view data relationship focuses on exploring the connection between the content of views, such as linking elements between views or coordinating between views (content is updated in a view when a user interacts with other views).
Content-aware layouts can be useful in specific analyses, such as investigating relationships, exploring patterns, filtering cross-views, finding similar items, or linking distinct items across views. 
In a scenario where a display space has many views, cross-view data relationships can help with the layout by separating/focusing on relevant views for analysis.
Additionally, the layout can be adjusted to focus on the cross-view data relationship to find similar or different items in different views, and then position the views according to this relationship for a more effective analysis.
We suggest and describe four content-aware layouts based on the two factors: view content and cross-view data relationship.

\vspace{-0.2cm}
\section{Layouts}
\rv{We define six perception-driven and three content-aware layouts as provided in Figure \ref{fig:layout} to serve as a guiding point for designers of MV systems. Furthermore, we discuss the characteristics of each layout as well as their usefulness.}
\vspace{-0.2cm}
\subsection{Perception-Driven Layout}
\textbf{Standard Layouts -}
The standard layouts adapted widely by most MV systems are flexible and fixed layouts. 
In the flexible layout (Figure \ref{fig:layout} (a)) designers provide no constraints on the views in the display space; there are no fixed positions, fixed sizes, or predefined salient views.
The user can easily drag, resize, and position views anywhere in the display space, which means MVs can be displayed in different sizes. 
Separate views can be positioned to overlap or appear side by side in this layout. 
Therefore, the user has a high-level of control over the views.
MV systems, such as Jigsaw \cite{jigsaw} and SimilarityExplorer \cite{similarity}, implemented flexible layouts.

Conversely, in a fixed layout (Figure \ref{fig:layout} (b)), designers assign a fixed position and size to all views in the display space. 
When a user opens or considers a view, it is displayed at its associated position and cannot be moved to another position or resized.
Separate views do not overlap in this type of layout, and views are usually positioned adjacently.
Designers can fix a salient view in an attractive location (such as the center of the display space), thereby directing users' attention to it. 
The user has minimal control of views with the layout predefined by the designer.
A fixed-layout has been designed and used in MV systems, such as BirdVis \cite{birdvis} and SwiftTuna \cite{tuna}.

\textbf{Cascade Layout -}
A cascade layout (Figure \ref{fig:layout} (c)) allows a designer to organize each view in a system into a stack of views known as a \emph{cascaded stack}.
The views in the stack are arranged so the user can observe portions of content in the view.
The user can focus on one or more specific views by selecting them from the cascaded stack, which maximizes the selected view.
This layout is similar to the cascade option on Windows 10; the layout allows for fixed or dynamic sizes, adjustable positions, and multiple salient views. 
The layout can accommodate multiple views within a display space with minimal clutter as separate views can be piled on top and overlapped. 
An exampled MV system implementing a cascade layout is \cite{heimerl2012visual}.

\textbf{Focused Layout -}
In the focused layout (Figure \ref{fig:layout} (d)), designers provide a \emph{focused window} at the center of the display space that allows users to focus or explore one or several specific views. 
A border surrounds the focused window to visualize views not present within it.
Depending on the display space, the views can be fixed or not in size, and their positions can be flexible. 
Current MV systems adopting this layout usually include supplemental views or controls on the side, often related to the salient view in the center. 
The designers use this layout when the user should focus on a particular view; some examples of MV systems are \cite{simon,zhao,bogl}.

\textbf{Split Layout -}
In the split layout (Figure \ref{fig:layout} (e)), the display space is split into one or more windows – a style used to organize several different views.
The user can split the display space into multiple separate windows, and views can be moved between the windows.
This layout is similar to the split-screen option available in Mac OS and Windows 10.
The views from the split windows can have different layouts, allowing for flexible or fixed sizes, positions, and multiple salient views. 
The split layout has been used by many MV systems \cite{van2014multivariate,palmas2014movexp}.

\textbf{Stacked Layout -}
In the stacked layout (Figure \ref{fig:layout} (f)), designers can arrange views on top of one another as they are sequentially displayed in the viewing space.
The position of the views in this layout is flexible, although the size of the views is fixed with no specific salient view. 
This layout can be designed according to a row or column stacked structure in which the views are arranged in a row or column format, respectively. 
It allows flexibility to arrange and remove views as required. 
The user can prioritize the views, analyze them sequentially and perform step-by-step analysis efficiently. 
The stack data structure inspires this layout.
Examples of MV systems that are designed similar to this layout are \cite{hajizadeh2013supporting,maljovec2016rethinking}.

\textbf{Tab Layout -}
The tab layout (Figure \ref{fig:layout} (g)) includes views that are entirely minimized in the form of a tab such that only the name of the view is visible. 
This layout has flexible positions, flexible sizes, and multiple salient views.
When a user hovers over a tab, the view is shown. 
This allows the user to either maximize the view for a more focused view or remove it from the display space. 
Maximized or focused views can be minimized to form tabs.
This layout is influenced by the tab option available on most operating systems.
Some MV systems designed based on this layout are \cite{unger2012visual,relex}

\vspace{-0.2cm}
\subsection{Content-Aware}
\textbf{ \rv{Hierarchical} Layout -}
Designers can implement the \rv{hierarchical} layout (Figure \ref{fig:layout} (h)) to create a tree-like structure between views with a root view and its leaf views.
The design of the layout enables users to keep linked views together and focus on several connections simultaneously.
Views can be added to the display space multiple times, allowing them to become either the root view or the leaf view of a tree structure. 
The linking of items across different views and the relation of content present in views through these links is an essential consideration in this layout.
A user can keep track of the views explored in the analytical process, which can help a user retrace their steps.
The layout is based on the tree data structure and can be helpful in tasks that require linking of views, hierarchy-based tasks, or step-based tasks.
An example of an MV system designed based on this layout is GraphTrail\cite{dunne2012graphtrail}.

\textbf{Composite Layout -}
In the composite layout (Figure \ref{fig:layout} (i)), designers can provide a system that allows users to utilize as much space as possible by presenting content in vacant areas or by integrating views.
The user can incorporate content into vacant spaces caused by any overlap between views.
The user can also choose particular views and combine them to create a combination of the selected views. 
The layout is focused on using most of the space available to provide essential content in the same display space.
\rv{Designers may also choose to leave space between views to depict related views.}
Similar to the OnSet technique \cite{Sadana}, this layout is a result of considering common patterns among views.
Designers can use this layout to display MVs in small display spaces and assist in identifying commonalities among views.
The various strategies or patterns that designers can consider to implement this layout are given in \cite{Javed}.

\textbf{Categorized Layout -}
Designers can utilize the categorized layout (Figure \ref{fig:layout} (j)) to group similar or related views in the MV system.
Similarities between the views need to be considered for producing this layout. 
The similarity between views can be observed by similar visualization in views or through the presentation of different perspectives of the same elements.
The content of the views are categorized, and similar views are displayed together in a separate window.
This layout enables the user to focus on a specific category of related views and can be used for finding similar or dissimilar elements.
\cite{zhang2014visual} depicts an MV system utilizing this layout to categorize different visualizations into separate areas of the display space.
\vspace{-0.2cm}
\section{Use Case}

\begin{figure}
\centering\includegraphics[width=\columnwidth,height=5cm]{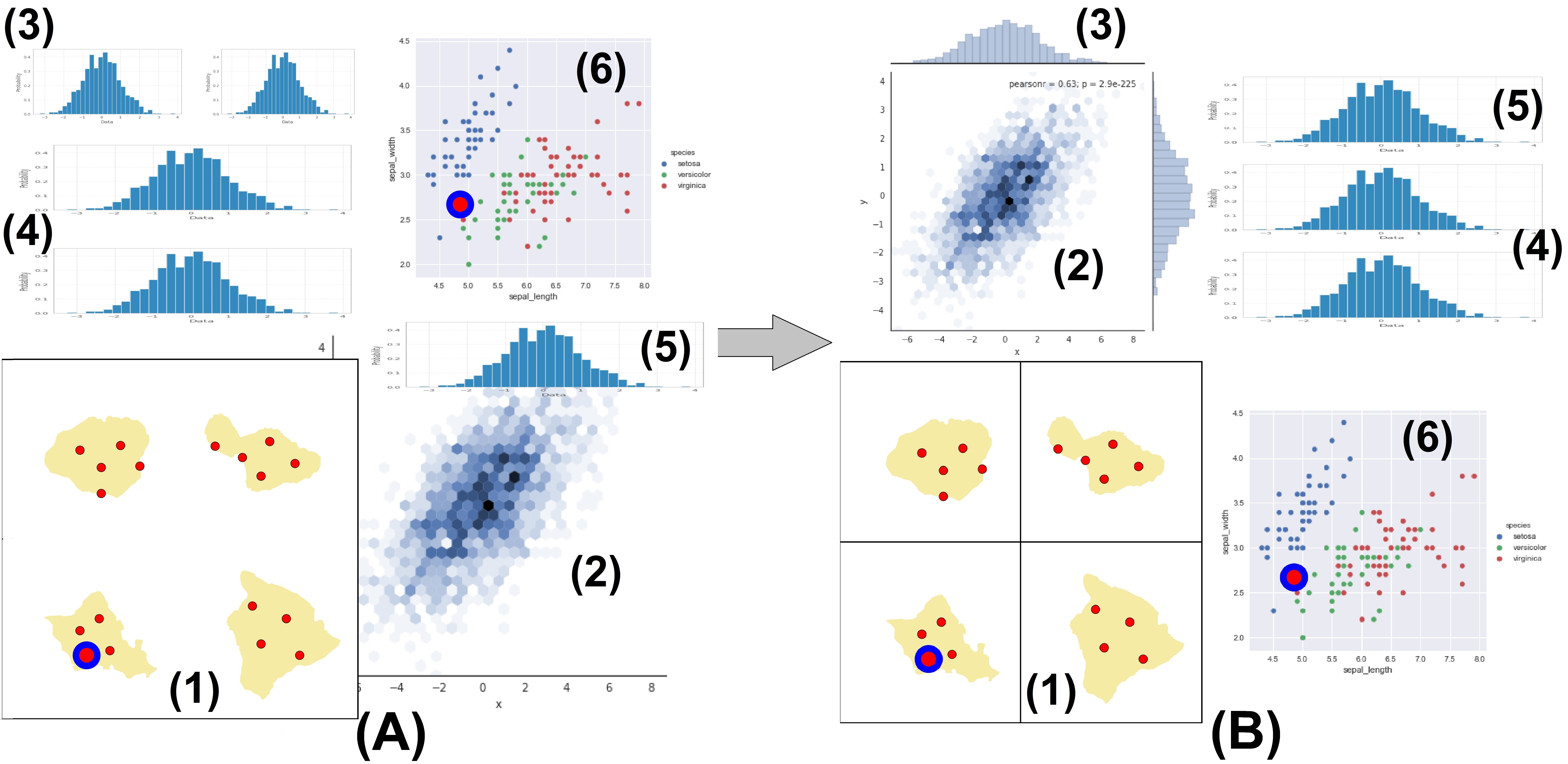}
\caption{An example of an MV system with a poor layout (A) and a revised, improved layout (B) using a systematic layout design.}
\label{fig:usecase}
\vspace{-1em}
\end{figure}

MV systems use multiple views to improve understanding of data, but most designers do not formally examine the relationships between the views (cross-view relationships) in a layout.
Neglecting the impacts on perception and understanding can lead to issues as shown in Figure~\ref{fig:usecase}(A).
The issues can be noticed by considering the factors related to perception and content.
The island visualizations shown in (1) are not separated, leading to potential confusion as to their relative geographic locations.
Here, views are not aligned, and the histogram view (5) blocks the visibility of some of the content (2).
Considering the \emph{separation} factor, the issue of overlap caused by (4) and the confusion in (1) is noticeable.
The \emph{relative position} of (5), (3), and (6) result in overlaps and disjointed views.
In addition, \emph{cross-view data relationship} is overlooked. 
Views (1) and (6) use linked highlighting (shown using the red dot), but they are not adjacent, meaning users may overlook the connection.
Finally, The histograms in (3) communicate the distribution of attributes that are also shown in the binned scatterplot in (2), but again they are not connected in the layout.

These issues can be addressed by implementing the suggested layouts in the MV system.
Implementing the categorized layout fixes the overlap issue by categorizing (4) and (5) in a separate area in the display space.
It also places views (1) and (6) side by side so that the user can easily identify the connection between the views; using the split layout, there is a separation in (1) to distinguish each visualization in the view.
In addition, we combine views (2) and (3) into a joint plot to better communicate the relationship between attributes by utilizing the composite layout.
The layout of the MV system is now improved (see Figure \ref{fig:usecase} (B)), conveying the information effectively in a small display space.
This shows the transition of an MV system from a poor layout to an improved layout, considering the factors and layouts proposed in this work.
Additionally, it demonstrates that a designer can integrate designs from one or more layouts when designing an MV system.

\vspace{-0.2cm}
\section{Discussion and Future Work}
Current MV systems are designed in a way that usually requires manual or semi-directed management of view layout.
By considering the content of views and their cross-view data relationships, it is possible to improve guidance for the design of MV layout. 
The factors and suggested MV layouts can also facilitate the transition from a user-directed layout to a steered, content-aware layout management that incorporates layout as a type of information synthesis. 

This paper investigates the challenges of layout design for MV systems and suggests factors to consider when designing MV systems. 
We do believe that this framework is effective in addressing challenges in layout design. However, our approach has some limitations; we do not claim to categorize \emph{all} layouts or factors, nor do we rank them. 
\rv{Some design factors for future study may include links between views, the number of related elements, and the semantic relevance of views.}
Using a proposed layout or optimizing a layout for a particular factor may not lead to an improved MV layout.
We believe that these findings can stimulate more research on the importance of layout design in MVs, taking into account view content and cross-view data relationships. 
Our future research will explore users' real-time cognitive performance based on their ability to analyze datasets using the design layouts described. 
We intend to develop more content-driven design layouts to support complex MV systems.
To further validate the layouts and factors, we will implement layouts similar to those shown in Figure~\ref{fig:usecase} and study user performance on analytical tasks.


\acknowledgments{
This research is supported in part by the NSF Grants IIS-2002082 and SMA-2022443.}

\bibliographystyle{abbrv-doi}

\bibliography{paper}
\end{document}